\documentclass[aps,prb,twocolumn,showpacs]{revtex4}
\usepackage{graphicx}
\begin{document}

\title{Spin dynamics and antiferromagnetic order in 
PrBa$_{2}$Cu$_{4}$O$_{8}$ studied by
Cu nuclear resonance}
\author{S.~Fujiyama}
\altaffiliation{Present address: Institute for Molecular Science, Okazaki, Aichi 444-8585, Japan}
\email{fujiyama@ims.ac.jp}
\author{M.~Takigawa}
\email{masashi@issp.u-tokyo.ac.jp}
\affiliation{The Institute for Solid State Physics, University of Tokyo, 
Kashiwa, Chiba 277-8581, Japan} 
\author{T.~Suzuki}
\author{N.~Yamada}
\affiliation{Department of Applied Physics and Chemistry, University of 
Electro-Communications, Chofu, Tokyo 182-8585, Japan} 
\author{S.~Horii}
\affiliation{Department of Applied Chemistry, 
University of Tokyo, Bunkyou-ku, 113-0033 Tokyo, Japan}
\author{Y.~Yamada}
\affiliation{Interdisciplinary Falulty of Science and Engineering,
Shimane University, Matsue 690-8504, Japan}

\date{\today}

\begin{abstract}
Results of the nuclear resonance experiments for the planar Cu sites in 
PrBa$_{2}$Cu$_{4}$O$_{8}$ are presented. The NMR spectrum at 1.5~K 
in zero magnetic field revealed an internal field 
of 6.1~T, providing evidence for an antiferromagnetic order of the planar Cu spins. 
This confirms that the CuO$_2$ planes are insulating, 
therefore, the metallic conduction in this material is entirely due 
to the one-dimensional zigzag Cu$_2$O$_2$ chains. The results of the 
spin-lattice relaxation rates measured by zero field NQR above 245 K in 
the paramagnetic state are explained by the theory for a 
Heisenberg model on a square lattice. 
\end{abstract}
\maketitle
%\twocolumn
%\tighten
\narrowtext

There has been increasing interest in 
quasi-one-dimensional correlated electrons. Theoretical studies on 
generalized Hubbard or \textit{t-J} models for chains and ladders have revealed a 
rich phase diagram associated with various instabilities towards spin 
density wave, charge order, and 
superconductivity~\cite{Schulz1994,Giamarchi1997,Dagotto1994}. Experimental 
studies have been focused mostly on organic and cuprate materials. Many 
phases including superconducting states were found, for example, in the 
organic Bechgaard salts~\cite{Gruner1988} and cuprate ladder 
materials~\cite{Uehara1996}.

Pr based cuprates also provide model systems of strongly correlated 
quasi-one-dimensional electrons. The best studied compound is 
PrBa$_{2}$Cu$_{3}$O$_{7}$, which is isostructural to the superconducting 
YBa$_{2}$Cu$_{3}$O$_{7}$ but yet insulating. The antiferromagnetic order 
in the CuO$_2$ planes found by NMR experiments~\cite{Reyes1990} and the 
large charge transfer energy gap in the optical conductivity spectrum 
along the $a$-direction~\cite{Takenaka1992} in PrBa$_{2}$Cu$_{3}$O$_{7}$ 
indicate that the electronic structure of the CuO$_2$ planes is similar to 
that in undoped Mott insulators such as YBa$_{2}$Cu$_{3}$O$_{6}$. 
Fehrenbacher and Rice proposed a model including localized hole states 
made of Pr-$4f$ and O-$2p_{\pi}$ hybridized orbitals~\cite{Fehren1993} to 
explain the insulating nature of the CuO$_2$ planes. The CuO$_2$ chains, 
on the contrary, remain paramagnetic and show mid-infrared optical 
peak~\cite{Takenaka1992}, consistent with the hole concentration of 
approximately 0.5 per Cu. Such a quarter-filled one dimensional band was 
also observed by angle resolved photoemission 
experiments~\cite{Mizokawa1999}. In the NMR/NQR experiments for the chain 
Cu sites, Gr\'{e}vin {\it et al.} found large enhancement of the 
spin-lattice and spin-spin relaxation rates due to electric quadrupole 
interaction~\cite{Grevin1998}. This suggests that the insulating behavior 
of the chains is due to charge ordering (charge density wave) instability. 

PrBa$_{2}$Cu$_{4}$O$_{8}$ has the same structure as the superconducting 
YBa$_{2}$Cu$_{4}$O$_{8}$ with Cu$_2$O$_2$ zigzag chains. This compound is 
not superconducting but remains metallic down to low 
temperatures~\cite{Yamada1994,Terasaki1996}. Muon spin relaxation 
experiments at zero field by Yamada {\it et al.} revealed the onset of an 
internal field at 220 K and a second magnetic transition at 17 K, which 
were ascribed to the order of Cu spins and Pr moments, 
respectively~\cite{Yamada1996}. Kikuchi {\it et al.} performed Cu NMR 
experiments on powder sample with the \textit{c}-axis aligned by magnetic 
field~\cite{Kikuchi1996}. They found that the signal from the planar Cu 
sites disappears below 250 K, which they attributed to the magnetic order of the 
planar Cu spins, while the signal from chain Cu sites remain intact. 
Recent resistivity measurements on single crystals by Horii {\it et al.}~\cite{Horii2000}
and by Hussey {\it et al.}~\cite{Hussey021}
revealed large in-plane anisotropy, indicating that the conduction is dominantly due 
to chains. Thus PrBa$_{2}$Cu$_{4}$O$_{8}$ provides unique 
example of metallic quasi-1D electrons without apparent disorder.

In this paper, we report the results of nuclear resonance experiments on the 
planar Cu sites in PrBa$_{2}$Cu$_{4}$O$_{8}$ at zero magnetic field. The results on 
the chain Cu sites are reported in a separate paper. The spectrum at 1.5~K 
revealed an internal field of 6.1~T due to antiferromagnetic order of 
the planar Cu spins. The nuclear spin-lattice relaxation rate $1/T_{1}$ 
was measured using the nuclear quadrupole resonance (NQR) signal at zero 
magnetic field in the paramagnetic state above 245~K. The temperature 
dependence of $1/T_{1}$ is consistent with the theoretical formula, $1/T_{1} 
\propto T^{1.5} \exp (1.13J/T)$, for the spin 1/2 Heisenberg model on a 
square lattice~\cite{Chakravarty1990} with the exchange integral $J$ = 1230~K 

The powder sample of PrBa$_{2}$Cu$_{4}$O$_{8}$ was synthesized by 
solid state reaction under high pressure as described in 
Ref.~\onlinecite{Yamada1994}.  The NMR/NQR 
experiments were performed with the standard spin-echo pulse sequence 
combined with the inversion recovery method for $T_1$ measurements.  

\begin{figure}[t]
\includegraphics*[width=8cm]{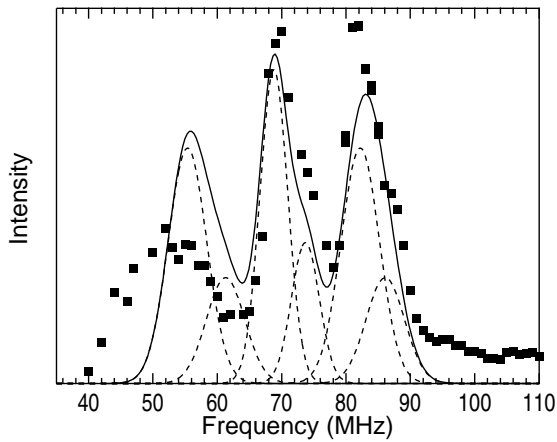}
\caption{\label{fig:AFNR}The spin-echo intensity for the planar Cu sites at zero magnetic 
field is plotted as a function of frequency at $T$=1.5 K (symbols). The solid 
line is fit to the spectrum as described in the text, which is the sum of six resonance 
lines shown by the dashed liines.}
\end{figure}

We show in Fig.~\ref{fig:AFNR} the spin-echo NMR spectrum at 1.5~K
in zero external field.  The spectrum has three distinct peaks at
52, 70 and 82~MHz.  Because the spin-echo decay time ($T_{2}$) is 
very short, we did not attempt to correct the spin echo intensity 
for possible frequency variation of $T_{2} $.
The spectrum with more than two peaks at frequencies much larger that
the typical NQR frequencies of high-$T_{c} $ cuprates ($\sim$ 30~MHz) 
points to the presence of a larger internal magnetic field due to antiferromagnetic order.

To understand the spectrum, we consider the Hamiltonian for spin 3/2 
nuclei in a magnetically ordered state in the presence of electric field 
gradient (EFG), which is written as,
\begin{equation}
{\cal H}=\gamma_{N}\hbar H_{\rm int}I_{z} +
\frac{eQ}{4I(2I-1)}V_{zz}(3I_{z}^{2}-I(I+1)). 
\label{eq:Hamiltonian}
\end{equation}
Here, $\gamma_{N}$ and $Q$ are the gyromagnetic ratio and the quadrupole 
moment of the nuclei, $H_{\rm int}$ is the magnitude of the internal magnetic 
field whose direction is taken to be the $z$-direction, and 
$V_{zz}=\partial ^2V/\partial z ^2$ is the $zz$-component of the EFG tensor. 
Since the analysis of the spectrum shown below indicates that the magnetic 
Zeeman interaction is much larger than the electric quadrupolar interaction, 
only the first order effect of the quadrupolar interaction is included in the 
above Hamiltonian. 

When the values of $H_{\rm int}$ and $V_{zz}$ are given for one isotope 
($^{63}$Cu), we can compute positions of all six resonance lines, three 
lines corresponding to the transitions $I_{z}=3/2 \leftrightarrow 1/2, 1/2 
\leftrightarrow -1/2, -1/2 \leftrightarrow -3/2$ for each of the two 
isotopes ($^{63}$Cu and $^{65}$Cu), by using the known ratios of the 
gyromagnetic ratios ($^{63}\gamma/^{65}\gamma =0.933$) and electric 
quadrupolar moments ($^{63}Q/^{65}Q=1.081$) for the two isotopes. The solid 
lines in Fig.~\ref{fig:AFNR} is the fit to the experimental spectrum 
with the values $H_{\rm int} = 6.1$~T and $\nu_{zz}=3eQV_{zz}/2hI(2I-1) = 
13.4$ MHz for $^{63}$Cu. The half width at half maximum for $^{63}$Cu is 
1.5~MHz for the central line ($I_{z}=1/2 \leftrightarrow -1/2$) and  
2.1~MHz for the satellite lines ($I_{z}=\pm 3/2 \leftrightarrow \pm 1/2$).  
Thus the NMR spectrum provides direct evidence for an antiferromagnetic 
order of the planer Cu spins. 

In the paramagnetic state above 245 K, we observed the nuclear quadrupole 
resonance (NQR) signal at zero field for $^{63}$Cu nuclei at $\nu_{Q}=31.2$~MHz, which  
is close to the resonance frequency in YBa$_{2}$Cu$_{3}$O$_{7}$ (31.2~MHz) and 
YBa$_{2}$Cu$_{4}$O$_{8}$ (29.7~MHz). Since in both compounds, EFG tensor 
is axially symmetric around the $c-$axis, we 
assume this to be the case in PrBa$_{2}$Cu$_{4}$O$_{8}$ as well. Then the 
angle $\theta$ between the $c$-axis and the magnetic hyperfine field in 
the ordered state is given by the relation
$\nu_{zz}=\nu_{Q}(3\cos^{2}\theta-1)/2$, leading to the values of 
$\theta=77^{\circ}$ or $38^{\circ}$. We adopt the former value since 
analysis of the NMR spectra in the antiferromagnetic state of both 
YBa$_{2}$Cu$_{3}$O$_{6}$~\cite{Yasuoka1988} and 
La$_{2}$CuO$_{4}$~\cite{Tsuda1988} concluded that the magnetic hyperfine 
field is parallel to the $ab-$plane within 10$^{\circ}$. This is also 
consistent with the fact that the ordered moment in the parent insulators 
of the high-$T_{c} $ cuprates lies approximately in the $ab-$plane. We note, 
however, the NMR spectrum in the ordered state in 
PrBa$_{2}$Cu$_{3}$O$_{7}$~\cite{Reyes1990,Matsumura1999} 
and PrBa$_{2}$Cu$_{3}$O$_{6}$~\cite{Yoshimura1994} show only one broad peak 
near 70 MHz with no resolved quadrupolar splitting, leading to a much 
smaller value of $\theta \sim 57^{\circ}$~\cite{Matsumura1999}. 

If we use the values of the hyperfine coupling constants in the Mila-Rice 
scheme~\cite{Mila1989} estimated for
YBa$_{2}$Cu$_{3}$O$_{7}$~\cite{Monien1991}, $A_{cc}$=-160, $A_{ab}$=34, 
$B$=40 kOe/$\mu_{B}$, where $A_{cc}$ and $A_{ab}$ is the hyperfine field 
from the onsite spin and $B$ is the supertransferred hyperfine field from a 
nearest neighbor spins, the magnitude and the direction of the ordered 
moment are estimated to be 0.50~$\mu_{B}$ and $4.6^{\circ}$ out of the 
$ab-$plane. The hyperfine coupling constants in PrBa$_{2}$Cu$_{4}$O$_{8}$, 
however, are not accurately known. The magnitude and the direction of the 
ordered moment thus obtained are subject to rather large uncertainties.  

In the paramagnetic state, we measured the nuclear spin lattice relaxation 
rate $1/T_{1}$ by using Cu NQR signal in the temperature range 245 - 
320 K. We could not observe NQR signal at lower temperatures.  This is  
consistent with the N\'{e}el temperature of 220~K revealed by both the 
disappearance of the planar Cu NMR signal~\cite{Kikuchi1996} at
a high magnetic field and the onset 
of the internal field at the muon sites~\cite{Yamada1996}. 

Nuclear relaxation in two-dimensional quantum antiferromagnets was 
theoretically studied by Chakravarty and Orbach~\cite{Chakravarty1990} in 
the renormalized classical regime, where the ground state is a N\'{e}el 
ordered state. For the case of spin-1/2 Heisenberg model on a 
square lattice with the nearest neighbor exchange $J$, $1/T_1$ is expressed as 
\begin{eqnarray}
\frac{1}{T_1} & = & \frac{0.134(2A_Q\gamma_N\hbar)^2}{2\pi\rho_s\hbar} 
\left(\frac{T}{2\pi\rho_s}\right)^{3/2}
\left(\frac{1}{1+T/2\pi\rho_s}\right)^3 \nonumber \\ & \times & 
\exp\left(\frac{2\pi\rho_s}{T}\right) , 
\label{eq:T1}
\end{eqnarray}
where $\rho_s=0.18J$ is the spin-stiffness constant and $A_Q=A_{ab}-4B$. 
When the temperature is much smaller than $2\pi\rho_s=1.13J$, 
$1/T_{1}T^{1.5}$ is expected to be proportional to $\exp(1.13J/T)$. 

\begin{figure}[t]
\includegraphics*[width=8cm]{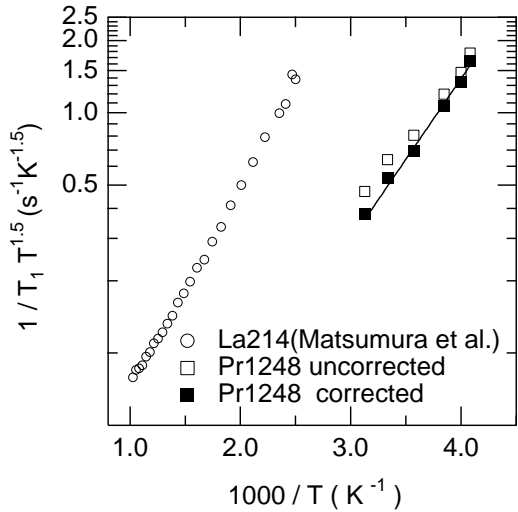}
\caption{\label{fig:T1}The nuclear relaxation rate in the paramagnetic state. 
The data of $1/T_{1}T^{1.5}$ are plotted against $1/T$. 
The raw experimental data for PrBa$_2$Cu$_4$O$_8$ are shown by open square. 
The results after subtracting the temperature independent contribution 
from Pr moments are shown by solid squares. For comparison, the data for 
La$_{2} $CuO$_{4}$ reported by Matsumura \textit{et al.}(Ref.~\onlinecite{Matsumura1994}) 
are also shown by open circles.}
\end{figure}

The observed $1/T_1$ at the planar Cu site includes the contribution 
from the Pr local moments through dipolar interaction, $(1/T_1)_{4f}$, 
in addition to the contribution from the planar Cu spins.  Thus we must 
subtract it from the experimental data before comparing with the 
theory. Since the temperature range of our measurements 
is higher than the ordering temperature of the Pr moments ($T_{N,4f}=17$~K), 
$(1/T_1)_{4f}$ can be estimated from the following formula valid in the high 
temperature limit~\cite{Moriya1956,Hammel1988},
\begin{equation}
(\frac{1}{T_{1} })_{4f} =\frac{2\sqrt 
{2\pi}}{3}\frac{\gamma_{N}^{2}(g_{J} \mu_{B} )^{2}}{\omega_{\rm{ex}}}
J(J+1)\Sigma_i \frac{3(5-3\cos ^{2} \theta_{j})}{4r_{j}^{6} }.
\label{eq:Moriya}
\end{equation} 
Where $J=4$, $g_J=0.8$ is the Land'{e}'s $g$-factor of Pr$^{3+}$ and $\theta_j$
is the angle between the $c$-axis and ${\bf r}_j$ connecting a Cu nucleus and
a neighboring Pr moment.  The exchange frequency $\omega_{\rm{ex}}$ is given as
\begin{equation}
\omega_{\rm ex} = \sqrt{\frac{2}{3}}\frac{J_{\rm 4f}}{\hbar}\sqrt{zS(S+1)} , 
\label{eq:omegaex}
\end{equation} 
where $S=1$, $z=4$ is the number of the nearest neighbor Pr sites,  
and $J_{\rm 4f}$ is the exchange interaction among Pr moments that can be 
estimated from the mean field expression for the ordering temperature,
$T_{N,4f} = J_{\rm 4f} z S(S+1)/3k_{B}$. 
We obtain $\omega_{\rm ex} = 1.94 \times 10^{12}$ [s$^{-1}$]. 
The sum over the Pr sites in Eq.~\ref{eq:Moriya} is dominated by the four nearest 
neighbors. By putting $\gamma_N=2\pi\times 1.1285 \times 10^{3} $[s$^{-1}$G$^{-1}$], 
$\cos ^2 \theta_j = 0.28$ and $\Sigma_{j}\frac{1}{r^{6}_{j}}=3.29\times 10^{45}$[cm$^{-6}$], 
we obtained $(1/T_1)_{4f}=550$[s$^{-1}$]. 

In Fig.~\ref{fig:T1}, we plot $1/T_{1}T^{1.5}$ against $1/T$ using the 
corrected data after subtracting $(1/T_1)_{4f}$ (solid squares).  As 
shown by the straight line, our results on PrBa$_{2}$Cu$_{4}$O$_{8}$ are 
compatible with the theoretical prediction. By fitting the data to 
Eq.~(\ref{eq:T1}), we obtained $J=1230$ K and $A_Q$ = 56 kG/$\mu_{B}$. Imai 
{\it et al.} first presented such an analysis for their experiments on 
La$_2$CuO$_4$~\cite{Imai1993}. They obtained $J= 1590$~K for
La$_2$CuO$_4$. The smaller $J$ for PrBa$_{2}$Cu$_{4}$O$_{8}$ is 
consistent with the lower N\'{e}el temperature. 

In summary, we confirmed that the ground state of PrBa$_{2}$Cu$_{4}$O$_{8}$ has 
an antiferromagnetic long range order of the planar Cu spins. From the 
analysis of the NMR spectrum at 1.5 K, the magnitude and the direction of 
the internal magnetic field are determined. The direction of the internal field 
is about ten degrees out of the $ab-$plane, indicating that the ordered 
moment lie approximately within the $ab$-plane, similarly to the case of 
YBa$_{2}$Cu$_{3}$O$_{6}$ and La$_{2}$CuO$_{4}$. This is, however, contrasting 
to the results reported for PrBa$_{2}$Cu$_{3}$O$_{7}$. 
The nuclear relaxation rate above 245 K follows 
the relation $1/T_{1}T^{1.5}\propto 1/T$, which is expected for 
two-dimensional spin-1/2 Heisenberg antiferromagnets on a square lattice. 

%Acknowlegement
We acknowledge Dr. Y. Itoh for providing us the data of $1/T_{1}$ in 
La$_{2}$CuO$_{4}$. This work is supported by the Grant in Aid of the Ministry of
Education, Culture, Sports, Science and Technology.  S. F. was supported by 
Research Fellowship for Young Scientists from Japan Scoiety for Promotion of Science.

\end{document}